\documentclass[12pt]{article}
\usepackage{amsmath,amssymb,amsthm,amsxtra,overpic,bbm,bm,epsfig,ulem}
\usepackage{color}
\usepackage{cite}

\usepackage{hyperref}
\usepackage{url}

\def\thefootnote{\fnsymbol{footnote}}

\begin{document}

\vspace{0.2cm}

\begin{center}
{\large\bf Emergent large flavor mixing from canonical and
inverse seesaws?}
\end{center}

\vspace{0.2cm}

\begin{center}
{\bf Zhi-zhong Xing$^{1,2}$}
\footnote{E-mail: xingzz@ihep.ac.cn}
\\
{\small $^{1}$Institute of High Energy Physics and School of Physical Sciences, \\
University of Chinese Academy of Sciences, Beijing 100049, China \\
$^{2}$Center of High Energy Physics, Peking University, Beijing 100871, China}
\end{center}

\begin{abstract}
While the canonical seesaw mechanism provides a most natural {\it qualitative}
interpretation of tiny masses for the three active neutrinos, it offers no
explanation for their large flavor mixing effects. The latter can be
regarded as an {\it emergent} consequence of this mechanism, in which case
we are left with an intriguing {\it cross seesaw} framework in the mass
basis of all the six Majorana neutrinos. To lower the mass scales of heavy
neutrinos, one is motivated to invoke the inverse seesaw mechanism but has
to pay the price for a {\it fine-tuned} cancellation between its two sets
of new degrees of freedom, in which case the largeness of active flavor
mixing is an {\it emergent} phenomenon as well. A comparison between the
{\it approximate} seesaw relations in the flavor basis and those {\it exact}
ones in the mass basis is also made.
\end{abstract}

\newpage

\def\thefootnote{\arabic{footnote}}
\setcounter{footnote}{0}

\section{Motivation}

The canonical seesaw mechanism~\cite{Minkowski:1977sc,Yanagida:1979as,
GellMann:1980vs,Glashow:1979nm,Mohapatra:1979ia}, accompanied by the
almost costless thermal leptogenesis mechanism~\cite{Fukugita:1986hr},
has widely been accepted as a most natural and most economical extension
of the standard model (SM) of particle physics to interpret why the masses
of three active neutrinos are so tiny and why the primordial antibaryons
have disappeared from the Universe. It is also fully compatible with
the keystones of the SM effective field theory after the relevant heavy
degrees of freedom (i.e., the fields of heavy Majorana neutrinos) are
integrated out~\cite{Weinberg:1979sa}, and hence has a great chance to
be accommodated in an ultimate UV-complete theory beyond the SM.

A reasonable question, which has not been seriously raised before, is
whether the large flavor mixing effects as observed in a variety of
neutrino oscillation experiments are a {\it natural} consequence or
{\it emergent} property of the canonical seesaw mechanism.

Another burning question is associated with how to naturally lower
possible mass scales of the hypothetical heavy Majorana neutrinos to
make the seesaw mechanism directly testable in the TeV energy
regions~\cite{Xing:2009in}. In this regard the inverse seesaw
mechanism, a straightforward extension of the canonical seesaw
framework with three neutral SM gauge-singlet fermions
and one neutral gauge-singlet scalar field~\cite{Wyler:1982dd,
Mohapatra:1986bd,Xing:2009hx}, has attracted a lot of interest. But to
what extent is the flavor structure of this mechanism free from any
intolerable unnaturalness~\cite{Giudice:2008bi}?

The present work aims to answer these two related questions. Instead
of making use of the well known but ``approximate" seesaw formula in
the {\it flavor} basis of the canonical or inverse seesaw mechanism,
we are going to look into the seesaw's structural symmetry with the help
of the ``exact" seesaw relation in the {\it mass} basis of the relevant
fermion fields. This new angle of view is enlightening as all the
{\it physical} seesaw flavor parameters take up their positions.
We find that the observed flavor mixing pattern of three active
neutrinos should be an {\it emergent} consequence of the canonical seesaw
mechanism, in which case we are left with an intriguing {\it cross
seesaw} framework. It is also remarkable that the inverse seesaw mechanism
works under the condition of a {\it fine-tuned} cancellation between
its two sets of new degrees of freedom in which case the large active
flavor mixing effects are an {\it emergent} phenomenon as well, but this
kind of structural unnaturalness may be fairly tolerable from a
phenomenological point of view.

\section{Canonical seesaw}

Without loss of generality, we start with the canonical seesaw Lagrangian
in the diagonal basis of both the charged-lepton Yukawa coupling matrix
and the right-handed neutrino mass matrix:
\begin{eqnarray}
-{\cal L}^{}_{\rm ss} \hspace{-0.2cm} & = & \hspace{-0.2cm}
\overline{\ell^{}_{\rm L}} \hspace{0.05cm} \widehat{Y}^{}_l
H \hspace{0.02cm} l^{}_{\rm R} +
\overline{\ell^{}_{\rm L}} \hspace{0.05cm}
Y^{}_\nu \widetilde{H} N^{}_{\rm R}
+ \frac{1}{2} \hspace{0.05cm}
\overline{(N^{}_{\rm R})^c} \hspace{0.05cm} D^{}_{\rm R} N^{}_{\rm R}
+ {\rm h.c.}
\nonumber \\
\hspace{-0.2cm} & = & \hspace{-0.2cm}
\overline{l^{}_{\rm L}} \hspace{0.05cm} \widehat{Y}^{}_l \hspace{0.03cm}
l^{}_{\rm R} \phi^0 + \frac{1}{2} \hspace{0.05cm}
\overline{\big[\begin{matrix} \nu^{}_{\rm L} & (N^{}_{\rm R})^c\end{matrix}
\big]} \left(\begin{matrix} {\bf 0} & Y^{}_\nu \phi^{0*} \cr
Y^T_\nu \phi^{0*} & D^{}_{\rm R} \end{matrix}\right)
\left[\begin{matrix} (\nu^{}_{\rm L})^c \cr N^{}_{\rm R} \end{matrix}\right]
+ \overline{\nu^{}_{\rm L}} \hspace{0.05cm} \widehat{Y}^{}_l \hspace{0.03cm}
l^{}_{\rm R} \phi^+ - \overline{l^{}_{\rm L}} \hspace{0.05cm} Y^{}_\nu
N^{}_{\rm R} \phi^- + {\rm h.c.} \; , \hspace{0.5cm}
\label{1}
\end{eqnarray}
where $\ell^{}_{\rm L} = \left(\nu^{}_{\rm L} \hspace{0.24cm}
l^{}_{\rm L}\right)^T$ is an $\rm SU(2)^{}_{\rm L}$ lepton
doublet with $\nu^{}_{\rm L} = \left(\nu^{}_{e \rm L} \hspace{0.24cm}
\nu^{}_{\mu \rm L} \hspace{0.24cm} \nu^{}_{\tau \rm L}\right)^T$ and
$l^{}_{\rm L} = \left(l^{}_{e \rm L} \hspace{0.24cm} l^{}_{\mu \rm L}
\hspace{0.24cm} l^{}_{\tau \rm L}\right)^T$ being the column vectors of
left-handed lepton fields, $\widetilde{H} \equiv {\rm i} \sigma^{}_2 H^*$
with $H = \left(\phi^+ \hspace{0.24cm} \phi^0\right)^T$ standing for
the Higgs doublet of the SM, $(\nu^{}_{\rm L})^c$ and $(N^{}_{\rm R})^c$
denote the charge-conjugated counterparts of $\nu^{}_{\rm L}$ and
$N^{}_{\rm R}$, $l^{}_{\rm R} = \left(l^{}_{e \rm R} \hspace{0.24cm}
l^{}_{\mu \rm R} \hspace{0.24cm} l^{}_{\tau \rm R}\right)^T$
and $N^{}_{\rm R} = \left(N^{}_{e \rm R} \hspace{0.24cm}
N^{}_{\mu \rm R} \hspace{0.24cm} N^{}_{\tau \rm R}\right)^T$ are the
column vectors of right-handed charged-lepton and neutrino fields which
belong to the $\rm SU(2)^{}_{\rm L}$ singlets, $\widehat{Y}^{}_l$ is
the diagonal charged-lepton Yukawa coupling matrix, $Y^{}_\nu$
is the arbitrary Yukawa coupling matrix of massive neutrinos, and
$D^{}_{\rm R}$ is a diagonal Majorana mass matrix. Note that the neutral
and charged components of the Higgs doublet have the mass dimension and
act like the complex numbers. The $\rm SU(2)^{}_{\rm L} \times
U(1)^{}_{\rm Y}$ gauge symmetry of ${\cal L}^{}_{\rm ss}$ will be
spontaneously broken after the scalar fields acquire their vacuum
expectation values $\langle \phi^\pm\rangle = 0$ and
$\langle \phi^0\rangle = v/\sqrt{2}$ with $v = \left(\sqrt{2}
\hspace{0.08cm} G^{}_{\rm F} \right)^{-1/2} \simeq 246~{\rm GeV}$,
in which case one is left with the Dirac neutrino mass matrix
$M^{}_{\rm D} \equiv Y^{}_\nu \langle \phi^0\rangle$.

The $6 \times 6$ symmetric complex matrix in Eq.~(\ref{1}) can be
formally diagonalized with the help of the well known Autonne-Takagi
transformation~\cite{Xing:2023adc},
\begin{eqnarray}
\mathbb{U}^\dagger \left ( \begin{matrix} {\bf 0} & Y^{}_\nu \phi^{0*}
\cr Y^T_\nu \phi^{0*} & D^{}_{\rm R} \cr \end{matrix} \right )
\mathbb{U}^* = \left( \begin{matrix} D^{}_\nu & {\bf 0} \cr {\bf 0} &
D^{}_N \cr \end{matrix} \right) \; ,
\label{2}
\end{eqnarray}
in which $\mathbb{U}$ is a $6\times 6$ unitary matrix, and the
diagonal and real matrices $D^{}_\nu$ and $D^{}_N$ are defined as
$D^{}_\nu \equiv {\rm Diag}\big\{m^{}_1, m^{}_2, m^{}_3\big\}$ and
$D^{}_N \equiv {\rm Diag}\big\{M^{}_4, M^{}_5, M^{}_6 \big\}$ with
$m^{}_i$ (for $i = 1, 2, 3$) and $M^{}_j$ (for $j = 4, 5, 6$)
being the working (or true) masses of light and heavy Majorana
neutrinos before (or after) spontaneous gauge symmetry breaking.
A novel block parametrization of $\mathbb{U}$, which reflects the
most salient features of the seesaw dynamics, is of the
form~\cite{Xing:2007zj,Xing:2011ur,Xing:2020ijf}
\footnote{At this point it is worth mentioning that a different block
decomposition of $\mathbb{U}$, in which the primary interplay effects
between the active and sterile flavor sectors are simply described by
three rotation angles as inspired by the original Wigner
parametrization~\cite{Wigner:1968}, has recently been proposed by
Zhou~\cite{Zhou:2025qwk}. This new and exact prescription in the
mass basis, which can be referred to as the Wigner-Zhou parametrization,
offers an alternative way of fully expressing the seesaw flavor structure
(see also Ref.~\cite{Blennow:2011vn} for a similar but approximate
description of this kind).}
\begin{eqnarray}
\mathbb{U} \hspace{-0.2cm} & = & \hspace{-0.2cm}
\left( \begin{matrix} I & {\bf 0} \cr {\bf 0} & U^\prime_0 \cr
\end{matrix} \right)
\left( \begin{matrix} A & R \cr S & B \cr \end{matrix} \right)
\left( \begin{matrix} U^{}_0 & {\bf 0} \cr {\bf 0} & I \cr
\end{matrix} \right)
\nonumber \\
\hspace{-0.2cm} & = & \hspace{-0.2cm}
\Big\{O^{}_{56} O^{}_{46} O^{}_{45}\Big\}
\Big[ \Big(O^{}_{36} O^{}_{26} O^{}_{16}\Big)
\Big(O^{}_{35} O^{}_{25} O^{}_{15}\Big)
\Big(O^{}_{34} O^{}_{24} O^{}_{14}\Big)
\Big] \Big\{O^{}_{23} O^{}_{13} O^{}_{12}\Big\} \; , \hspace{0.5cm}
\label{3}
\end{eqnarray}
where the $3\times 3$ unitary matrices $U^{}_0$ and $U^\prime_0$
characterize the primary flavor mixing effects in the respective
active (light) and sterile (heavy) neutrino sectors, the $3\times 3$
matrices $A$, $B$, $R$ and $S$ describe some tiny interplays between the
active and sterile sectors via the Yukawa interactions of massive neutrinos,
and each of the fifteen $6\times 6$ Euler-like unitary matrices $O^{}_{ij}$
(for $1 \leq i < j \leq 6$) has the following properties: its $(i, i)$ and
$(j, j)$ entries are both identical to $c^{}_{ij} \equiv \cos\theta^{}_{ij}$
with $\theta^{}_{ij}$ being a flavor mixing angle and lying in the first
quadrant, its other four diagonal elements are all equal to one, its
$(i, j)$ and $(j, i)$ entries are given respectively by
$\hat{s}^{*}_{ij} \equiv e^{-{\rm i}\delta^{}_{ij}} \sin\theta^{}_{ij}$ and
$-\hat{s}^{}_{ij} \equiv -e^{{\rm i}\delta^{}_{ij}} \sin\theta^{}_{ij}$ with
$\delta^{}_{ij}$ being a CP-violating phase, and its other off-diagonal
elements are all equal to zero. In the mass basis of all the charged-lepton
and neutrino fields, the leptonic weak charged- and neutral-current
interactions turn out to be
\begin{eqnarray}
-{\cal L}^{}_{\rm cc} \hspace{-0.2cm} & = & \hspace{-0.2cm}
\frac{g}{\sqrt{2}} \hspace{0.1cm}
\overline{\big(\begin{matrix} e & \mu & \tau\end{matrix}\big)^{}_{\rm L}}
\hspace{0.1cm} \gamma^\mu \left[ U \left( \begin{matrix} \nu^{}_{1}
\cr \nu^{}_{2} \cr \nu^{}_{3} \cr\end{matrix}
\right)^{}_{\hspace{-0.08cm} \rm L}
+ R \left(\begin{matrix} N^{}_4 \cr N^{}_5 \cr N^{}_6
\cr\end{matrix}\right)^{}_{\hspace{-0.08cm} \rm L} \hspace{0.05cm} \right]
W^-_\mu + {\rm h.c.} \; ,
\nonumber \\
-{\cal L}^{}_{\rm nc} \hspace{-0.2cm} & = & \hspace{-0.2cm}
\frac{g}{2 \cos\theta^{}_{\rm W}} \left\{
\overline{\big(\begin{matrix} \nu^{}_1 & \nu^{}_2 & \nu^{}_3
\end{matrix}\big)^{}_{\rm L}}
\hspace{0.1cm} \gamma^\mu \left[ U^\dagger U \left( \begin{matrix} \nu^{}_{1}
\cr \nu^{}_{2} \cr \nu^{}_{3} \cr\end{matrix}
\right)^{}_{\hspace{-0.08cm} \rm L}
+ U^\dagger R \left(\begin{matrix} N^{}_4 \cr N^{}_5 \cr N^{}_6
\cr\end{matrix}\right)^{}_{\hspace{-0.08cm} \rm L} \hspace{0.05cm} \right]
Z^0_\mu \right. \hspace{0.4cm}
\nonumber \\
\hspace{-0.2cm} & & \hspace{-0.2cm}
+ \left. \overline{\big(\begin{matrix} N^{}_4 & N^{}_5 & N^{}_6
\end{matrix}\big)^{}_{\rm L}}
\hspace{0.1cm} \gamma^\mu \left[ R^\dagger U \left( \begin{matrix} \nu^{}_{1}
\cr \nu^{}_{2} \cr \nu^{}_{3} \cr\end{matrix}
\right)^{}_{\hspace{-0.08cm} \rm L}
+ R^\dagger R \left( \begin{matrix} N^{}_{4}
\cr N^{}_{5} \cr N^{}_{6} \cr\end{matrix}
\right)^{}_{\hspace{-0.08cm} \rm L} \right] Z^0_\mu \hspace{0.05cm} \right\} \; ,
\label{4}
\end{eqnarray}
in which $g$ is the weak gauge coupling constant,
$\theta^{}_{\rm W}$ denotes the Weinberg angle of weak interactions,
$U = A U^{}_0$ represents the $3 \times 3$ Pontecorvo-Maki-Nakagawa-Sakata
(PMNS) matrix of lepton flavor mixing~\cite{Pontecorvo:1957cp,Maki:1962mu,
Pontecorvo:1967fh}, $R$ signifies the strengths of Yukawa interactions
of massive Majorana neutrinos and satisfies the unitarity condition
$U U^\dagger + R R^\dagger = A A^\dagger + R R^\dagger = I$. One may find
the exact and explicit expressions of $U^{}_0$, $A$ and $R$ in terms of
the Euler-like rotation angles and CP-violating phases in Appendix A.
Note that the pattern of $A$ is a lower triangular matrix, and that of
$U^{}_0$ is essentially the same as the standard parametrization advocated
by the Particle Data Group~\cite{ParticleDataGroup:2024cfk}.

It is worth emphasizing that $U^{}_0 = A^{-1} U$ and $R$ are {\it inherently}
correlated with each other via the {\it exact} seesaw relationship in the
physical (mass) basis of all the six Majorana neutrinos
\footnote{The {\it approximate} seesaw formula
$M^{}_\nu \equiv U^{}_0 D^{}_\nu U^T_0 \simeq
\left({\rm i} M^{}_{\rm D}\right) M^{-1}_{\rm R}
\left({\rm i} M^{}_{\rm D}\right)^T$ is well known in the flavor basis.},
\begin{eqnarray}
U^{}_0 D^{}_{\nu} U^T_0 = \left({\rm i} A^{-1} R\right)
D^{}_{N} \left({\rm i} A^{-1} R\right)^T \; ,
\label{5}
\end{eqnarray}
which can simply be obtained from substituting Eq.~(\ref{3}) into
Eq.~(\ref{2}). It is evident that the two sides of this elegant equation
exhibits a striking structural symmetry between the light (active)
and heavy (sterile) flavor sectors, implying that the two sectors are
two sides of the same coin in some sense. Some comments are in order.
\begin{itemize}
\item     The eighteen {\it original} seesaw flavor parameters include
three heavy Majorana neutrino masses in $D^{}_N$, nine active-sterile
flavor mixing angles and six independent CP-violating phases in $R$
or $A$, as can be clearly seen from an explicit Euler-like parametrization
of $A$ and $R$~\cite{Xing:2007zj,Xing:2011ur,Xing:2020ijf}.
In comparison, the nine {\it derivational} seesaw flavor parameters
include three light Majorana neutrino masses in $D^{}_\nu$, three active
flavor mixing angles and three CP-violating phases in $U^{}_0$, which
can all be calculated in terms of those original flavor
parameters~\cite{Xing:2024xwb,Xing:2024gmy}.

\item     Non-unitarity of the PMNS matrix $U$ is characterized by the
deviation of $U U^\dagger = A A^\dagger$ from the identity matrix $I$,
and its upper bound is found to be at most of ${\cal O}(10^{-3})$ from
a global analysis of today's precision electroweak and flavor
measurements~\cite{Blennow:2023mqx,Xing:2024gmy}. That is why
$A \simeq I$, $U \simeq U^{}_0$ and $A^{-1} R \simeq R$ hold as very
good approximations, implying that current neutrino oscillation experiments
are still insensitive to small unitarity violation of $U$.

\item     As the three eigenvalues of $U^{}_0 D^{}_\nu U^T_0$ on the
left-hand side of Eq.~(\ref{5}) are just the diagonal elements of
$D^{}_\nu$ (i.e., $m^{}_i$ for $i = 1, 2, 3$) below
${\cal O}(0.1)$ eV~\cite{ParticleDataGroup:2024cfk}, the corresponding
eigenvalues of $\left({\rm i} A^{-1} R\right) D^{}_{N} \left({\rm i}
A^{-1} R\right)^T$ on the right-hand side of Eq.~(\ref{5}) must be
the same. This observation implies that the smallness of nine active-sterile
flavor mixing angles $\theta^{}_{ij}$ (for $i = 1, 2, 3$ and $j = 4, 5, 6$)
in $A^{-1} R$ is absolutely demanded to compensate for the largeness of
$M^{}_4$, $M^{}_5$ and $M^{}_6$ in $D^{}_N$, so as to make the seesaw
mechanism work. In other words, the smallness of $m^{}_i$ originates from
the very fact that the Yukawa interactions between the active and sterile
sectors of Majorana neutrinos are weak enough to suppress the largeness
of $M^{}_j$.
\end{itemize}
While the tiny eigenvalues of $\left({\rm i} A^{-1} R\right) D^{}_{N}
\left({\rm i} A^{-1} R\right)^T$ assure that $m^{}_1$, $m^{}_2$ and
$m^{}_3$ are extremely small as compared with the masses of all the
other fundamental fermions, they are unable to give any hints about the
pattern of $U^{}_0$ even in a qualitative way.

The explicit flavor mixing pattern of three active neutrinos must be
sensitive to how off-diagonal the texture of
$\left({\rm i} A^{-1} R\right) D^{}_{N} \left({\rm i} A^{-1} R\right)^T$
can be in a given seesaw model, and hence the general seesaw mechanism
itself offers no explanation of why the three flavor mixing
angles of $U^{}_0$ are large --- at least remarkably larger than
their counterparts in the same parametrization of the $3\times 3$
Cabibbo-Kobayashi-Maskawa (CKM) quark flavor mixing
matrix~\cite{ParticleDataGroup:2024cfk} --- as observed
in a variety of neutrino oscillation experiments. But one may still see
how the exact seesaw relation in Eq.~(\ref{5}) transmits a potential flavor
mixing pattern hidden in the Yukawa interactions of Majorana
neutrinos to the active neutrino sector.

The point is that Eq.~(\ref{5}) reveals a kind of symmetry between the
flavor textures of active (light) and sterile (heavy) Majorana neutrinos,
so an underlying flavor symmetry of the Yukawa coupling
matrix $Y^{}_\nu \propto R D^{}_N/\langle \phi^0\rangle$ may naturally
manifest itself in the effective light Majorana neutrino mass matrix
$M^{}_\nu \equiv U^{}_0 D^{}_\nu U^T_0$. For the sake of illustration, let
us consider a special but phenomenologically favored lepton flavor mixing
pattern of the form~\cite{Xing:2006ms}
\begin{eqnarray}
U^{}_0 \hspace{-0.2cm} & = & \hspace{-0.2cm}
\left( \begin{matrix} \displaystyle\sqrt{\frac{2}{3}}
& \displaystyle\sqrt{\frac{1}{3}} \hspace{0.1cm} c^{}_\theta
& \displaystyle {\rm i} \sqrt{\frac{1}{3}} \hspace{0.1cm} s^{}_\theta \cr
\vspace{-0.35cm} \cr
\displaystyle -\sqrt{\frac{1}{6}}
& \displaystyle ~\sqrt{\frac{1}{3}} \hspace{0.1cm} c^{}_\theta +
{\rm i} \sqrt{\frac{1}{2}} \hspace{0.1cm} s^{}_\theta ~
& \displaystyle +\sqrt{\frac{1}{2}} \hspace{0.1cm} c^{}_\theta +
{\rm i} \sqrt{\frac{1}{3}} \hspace{0.1cm} s^{}_\theta \cr
\vspace{-0.35cm} \cr
\displaystyle -\sqrt{\frac{1}{6}}
& \displaystyle \sqrt{\frac{1}{3}} \hspace{0.1cm} c^{}_\theta -
{\rm i} \sqrt{\frac{1}{2}} \hspace{0.1cm} s^{}_\theta
& \displaystyle -\sqrt{\frac{1}{2}} \hspace{0.1cm} c^{}_\theta +
{\rm i} \sqrt{\frac{1}{3}} \hspace{0.1cm} s^{}_\theta \cr
\end{matrix} \right) \; , \;\;\;\;\;
\label{6}
\end{eqnarray}
where $c^{}_\theta \equiv \cos\theta$ and $s^{}_\theta \equiv \sin\theta$
with $\theta$ being a small angle which is adjustable to fit current
neutrino oscillation data. In this case one may easily arrive at the
relations
\begin{eqnarray}
\left(M^{}_\nu\right)^{}_{e \tau} =
\left(M^{}_\nu\right)^*_{e \mu} \; , \quad
\left(M^{}_\nu\right)^{}_{\tau \tau} =
\left(M^{}_\nu\right)^*_{\mu \mu} \; ,
\label{7}
\end{eqnarray}
corresponding to the popular $\mu$-$\tau$ reflection
symmetry~\cite{Harrison:2002et,Xing:2015fdg,Xing:2022uax}. Then the exact
seesaw formula in Eq.~(\ref{5}) guarantees the relations
\begin{eqnarray}
\left[\left({\rm i} A^{-1} R\right) D^{}_{N} \left({\rm i}
A^{-1} R\right)^T\right]^{}_{e \tau}
\hspace{-0.2cm} & = & \hspace{-0.2cm}
\left[\left({\rm i} A^{-1} R\right) D^{}_{N} \left({\rm i}
A^{-1} R\right)^T\right]^*_{e \mu} \; ,
\nonumber \\
\left[\left({\rm i} A^{-1} R\right) D^{}_{N} \left({\rm i}
A^{-1} R\right)^T\right]^{}_{\tau \tau}
\hspace{-0.2cm} & = & \hspace{-0.2cm}
\left[\left({\rm i} A^{-1} R\right) D^{}_{N} \left({\rm i}
A^{-1} R\right)^T\right]^*_{\mu \mu} \; . \hspace{0.6cm}
\label{8}
\end{eqnarray}
Given $A \simeq I$, these two equalities mean that a flavor symmetry
imposed on the Yukawa coupling matrix $Y^{}_\nu$ in the basis of diagonal
$\widehat{Y}^{}_l$ and $D^{}_N$ can help a lot to shape the flavor texture
of three active neutrinos~\cite{Xing:2022oob}. One may therefore argue
that such flavor symmetries could be a dynamical reason for the {\it emergence}
of large flavor mixing in the light Majorana neutrino sector, making the
phenomena of neutrino oscillations experimentally observable.

\begin{figure}[t]
\begin{center}
\includegraphics[width=12.5cm]{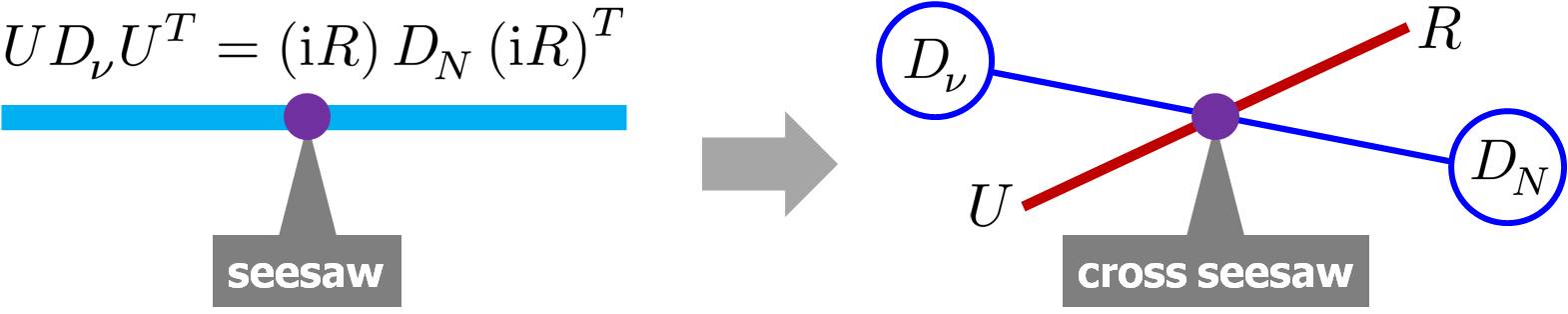}
\caption{A schematic illustration of the exact seesaw relation
$U D^{}_{\nu} U^T = \left({\rm i} R\right) D^{}_{N} \left({\rm i}
R\right)^T$, which exhibits a structural symmetry between light and heavy
Majorana neutrinos, and of its cross seesaw scenario with large flavor
mixing as an emergent phenomenon.}
\label{fig1}
\end{center}
\end{figure}

To avoid any model-dependent details in this connection, let us simply require
that $M^{}_j \gg \langle \phi^0\rangle$ (for $j = 4, 5, 6$) hold and there
be no severely contrived structural cancellation on the right-hand side of
Eq.~(\ref{5})~\cite{Kersten:2007vk} --- two bighearted criteria to assure
naturalness of the canonical seesaw mechanism. Then there must be
a huge hierarchy of at least more than thirteen orders of magnitude
between the light and heavy Majorana neutrino masses, in which case none of
the nine active-sterile flavor mixing angles in $R$ can be larger than
${\cal O}(10^{-6})$ (i.e., these angles are considerably smaller than the
observed values of three active flavor mixing angles). So treating the
seesaw framework as a {\it cross seesaw} system should be quite reasonable,
as illustrated in Fig.~\ref{fig1}, although the {\it emergent} pattern of $U$
remains a puzzle from the theoretical point of view. All in all, the
intrinsic properties of a general seesaw mechanism are far more complicated
than usually expected.

\section{Inverse seesaw}

The inverse seesaw mechanism is known as a relatively natural extension
of the canonical seesaw framework to lower the mass scales of heavy
Majorana neutrinos to the TeV energy regions. Besides the left- and
right-handed neutrino fields $\nu^{}_{\alpha {\rm L}}$ and
$N^{}_{\alpha {\rm R}}$ (for $\alpha = e, \mu, \tau$), the inverse seesaw
contains three neutral SM gauge-singlet fermions $S^{}_{\alpha {\rm R}}$
(for $\alpha = e, \mu, \tau$) and a scalar singlet $\Phi$. Its
gauge-invariant lepton mass terms in the basis of both the diagonal
charged-lepton Yukawa coupling matrix and the diagonal extra neutral fermion
mass matrix include
\begin{eqnarray}
-{\cal L}^{\prime}_{\rm ss} \hspace{-0.2cm} & = & \hspace{-0.2cm}
\overline{\ell^{}_{\rm L}} \hspace{0.05cm} \widehat{Y}^{}_l
H \hspace{0.02cm} l^{}_{\rm R} +
\overline{\ell^{}_{\rm L}} \hspace{0.05cm}
Y^{}_\nu \widetilde{H} N^{}_{\rm R} + \overline{(N^{}_{\rm R})^c} \hspace{0.05cm}
Y^{}_S \Phi S^{}_{\rm R} + \frac{1}{2} \overline{(S^{}_{\rm R})^c} \hspace{0.05cm}
\hat{\mu} S^{}_{\rm R} + {\rm h.c.}
\nonumber \\
\hspace{-0.2cm} & = & \hspace{-0.2cm}
\overline{l^{}_{\rm L}} \hspace{0.05cm} \widehat{Y}^{}_l \hspace{0.03cm}
l^{}_{\rm R} \phi^0 +
\frac{1}{2} \hspace{0.05cm}
\overline{\big[\begin{matrix} \nu^{}_{\rm L} & (N^{}_{\rm R})^c
& (S^{}_{\rm R})^c\end{matrix}\big]}
\left(\begin{matrix} {\bf 0} & Y^{}_\nu \phi^{0*} & {\bf 0} \cr
Y^T_\nu \phi^{0*} & {\bf 0} & Y^{}_S \Phi \cr
{\bf 0} &  Y^T_S \Phi & \hat{\mu} \end{matrix}\right)
\left[\begin{matrix} (\nu^{}_{\rm L})^c \cr N^{}_{\rm R} \cr S^{}_{\rm R}
\end{matrix}\right]
\nonumber \\
\hspace{-0.2cm} & & \hspace{-0.2cm}
+ \hspace{0.08cm} \overline{\nu^{}_{\rm L}} \hspace{0.05cm} \widehat{Y}^{}_l
\hspace{0.03cm} l^{}_{\rm R} \phi^+
- \overline{l^{}_{\rm L}} \hspace{0.05cm} Y^{}_\nu N^{}_{\rm R} \phi^-
+ {\rm h.c.} \; , \hspace{0.6cm}
\label{9}
\end{eqnarray}
where $S^{}_{\rm R} = (\nu^{}_{e \rm R} \hspace{0.24cm} \nu^{}_{\mu \rm R}
\hspace{0.24cm} \nu^{}_{\tau \rm R})^T$ is the vector column of three
gauge-singlet fermion fields, $\hat{\mu}$ denotes a diagonal mass matrix, and
the other notations are self-explanatory. After spontaneous symmetry breaking,
one is left with nonzero $\langle\phi^0\rangle$ and $\langle \Phi\rangle$
together with $\langle\phi^\pm\rangle = 0$, and then obtains the mass matrices
$M^{}_{\rm D} \equiv Y^{}_\nu \langle \phi^0\rangle$ and
$M^{}_S \equiv Y^{}_S \langle \Phi\rangle$. The $9 \times 9$
complex matrix in Eq.~(\ref{9}) can be diagonalized via the following
Autonne-Takagi transformation:
\begin{eqnarray}
\mathbb{U}^{\prime\dagger}
\left(\begin{matrix} {\bf 0} & Y^{}_\nu \phi^{0*} & {\bf 0} \cr
Y^T_\nu \phi^{0*} & {\bf 0} & Y^{}_S \Phi \cr
{\bf 0} &  Y^T_S \Phi & \hat{\mu} \end{matrix}\right) \mathbb{U}^{\prime *}
= \left( \begin{matrix} D^{}_{\nu} & {\bf 0} & {\bf 0} \cr
{\bf 0} & D^{}_N & {\bf 0} \cr
{\bf 0} & {\bf 0} & D^{}_S \end{matrix} \right) \; ,
\label{10}
\end{eqnarray}
where $D^{}_\nu = \{m^{}_1, m^{}_2, m^{}_3\}$,
$D^{}_N = \{M^{}_4, M^{}_5, M^{}_6\}$ and
$D^{}_S = \{M^{\prime}_7, M^{\prime}_8, M^{\prime}_9\}$ with $m^{}_i$,
$M^{}_j$ and $M^{\prime}_{j^\prime}$ being the respective working masses of
three active neutrinos $\nu^{}_i$, three sterile neutrinos $N^{}_j$ and three
extra neutral fermions $N^{\prime}_{j^\prime}$ (for $i = 1, 2, 3$;
$j = 4, 5, 6$; and $j^\prime = 7, 8, 9$). Like Eq.~(\ref{3}), a block
decomposition of the $9 \times 9$ unitary matrix $\mathbb{U}^\prime$ of
the form~\cite{Han:2021qum}
\begin{eqnarray}
\mathbb{U}^\prime \hspace{-0.2cm} & = & \hspace{-0.2cm}
\underline{\left( \begin{matrix} I & 0 & 0 \cr 0 & I & 0 \cr 0 & 0 & S^{}_0
\end{matrix} \right)}
\left( \begin{matrix} I & 0 & 0 \cr 0 & A^{}_3 & R^{}_3 \cr 0 & S^{}_3 & B^{}_3
\end{matrix} \right)
\underline{\left( \begin{matrix} I & 0 & 0 \cr 0 & U^{\prime}_0 & 0 \cr 0 & 0 & I
\end{matrix} \right)}
\left( \begin{matrix} A^{}_2 & 0 & R^{}_2 \cr 0 & I & 0 \cr S^{}_2 & 0 & B^{}_2
\end{matrix} \right)
\left( \begin{matrix} A^{}_1 & R^{}_1 & 0 \cr S^{}_1 & B^{}_1 & 0 \cr 0 & 0 & I
\end{matrix} \right)
\underline{\left( \begin{matrix} U^{}_0 & 0 & 0 \cr 0 & I & 0 \cr 0 & 0 & I
\end{matrix} \right)} \hspace{0.6cm}
\nonumber \\
\hspace{-0.2cm} & = & \hspace{-0.2cm}
\Big\{O^{}_{89} O^{}_{79} O^{}_{78}\Big\}
\Big[ \Big(O^{}_{69} O^{}_{59} O^{}_{49}\Big)
\Big(O^{}_{68} O^{}_{58} O^{}_{48}\Big)
\Big(O^{}_{67} O^{}_{57} O^{}_{47}\Big)\Big]
\Big\{O^{}_{56} O^{}_{46} O^{}_{45}\Big\}
\Big[ \Big(O^{}_{39} O^{}_{29} O^{}_{19}\Big)
\nonumber \\
\hspace{-0.2cm} & & \hspace{-0.2cm}
\Big(O^{}_{38} O^{}_{28} O^{}_{18}\Big)
\Big(O^{}_{37} O^{}_{27} O^{}_{17}\Big)\Big]
\Big[ \Big(O^{}_{36} O^{}_{26} O^{}_{16}\Big)
\Big(O^{}_{35} O^{}_{25} O^{}_{15}\Big)
\Big(O^{}_{34} O^{}_{24} O^{}_{14}\Big)\Big]
\Big\{O^{}_{23} O^{}_{13} O^{}_{12}\Big\} \;\;
\label{11}
\end{eqnarray}
is particularly instructive as it helps reflect the three (underlined)
primary flavor mixing sectors of active neutrinos ($U^{}_0$), sterile
neutrinos ($U^\prime_0$) and extra neutral fermions ($S^{}_0$), together
with the interplays between any two of the three sectors ($A^{}_i$,
$B^{}_i$, $R^{}_i$ and $S^{}_i$ for $i = 1, 2, 3$). Each of the
thirty-six $9\times 9$ Euler-like unitary matrices $O^{}_{ij}$
(for $1 \leq i < j \leq 9$) has the following properties: its $(i, i)$ and
$(j, j)$ entries are both identical to $c^{}_{ij} \equiv \cos\theta^{}_{ij}$
with $\theta^{}_{ij}$ being a flavor mixing angle and lying in the first
quadrant, its other seven diagonal elements are all equal to one, its
$(i, j)$ and $(j, i)$ entries are given respectively by
$\hat{s}^{*}_{ij} \equiv e^{-{\rm i}\delta^{}_{ij}} \sin\theta^{}_{ij}$ and
$-\hat{s}^{}_{ij} \equiv -e^{{\rm i}\delta^{}_{ij}} \sin\theta^{}_{ij}$ with
$\delta^{}_{ij}$ being a CP-violating phase, and its other off-diagonal
elements are all vanishing. The expressions of $A^{}_1$, $R^{}_1$ and
$U^{}_0$ are exactly the same as those of $A$, $R$ and $U^{}_0$ shown in
Appendix A, and the analytical results of $(A^{}_2, R^{}_2, U^{\prime}_0)$
and $(A^{}_3, R^{}_3, S^{}_0)$ are exactly parallel to those of
$(A^{}_1, R^{}_1, U^{}_0)$ with a replacement of the corresponding flavor
mixing angles and CP-violating phases~\cite{Han:2021qum}.

In the mass basis of all the relevant lepton fields, the leptonic weak charged-
and neutral-current interactions can be expressed as
\begin{eqnarray}
-{\cal L}^{\prime}_{\rm cc} \hspace{-0.2cm} & = & \hspace{-0.2cm}
\frac{g}{\sqrt{2}} \hspace{0.1cm}
\overline{\big(\begin{matrix} e & \mu & \tau\end{matrix}\big)^{}_{\rm L}}
\hspace{0.1cm} \gamma^\mu \left[ U \left( \begin{matrix} \nu^{}_{1}
\cr \nu^{}_{2} \cr \nu^{}_{3} \cr\end{matrix}
\right)^{}_{\hspace{-0.08cm} \rm L}
+ R \left(\begin{matrix} N^{}_4 \cr N^{}_5 \cr N^{}_6
\cr\end{matrix}\right)^{}_{\hspace{-0.08cm} \rm L}
+ R^\prime \left(\begin{matrix} N^{\prime}_7 \cr N^{\prime}_8 \cr
N^{\prime}_9 \cr\end{matrix}\right)^{}_{\hspace{-0.08cm} \rm L}
\hspace{0.05cm} \right] W^-_\mu + {\rm h.c.} \; ,
\nonumber \\
-{\cal L}^{\prime}_{\rm nc} \hspace{-0.2cm} & = & \hspace{-0.2cm}
\frac{g}{2 \cos\theta^{}_{\rm W}} \left\{
\overline{\big(\begin{matrix} \nu^{}_1 & \nu^{}_2 & \nu^{}_3
\end{matrix}\big)^{}_{\rm L}}
\hspace{0.1cm} \gamma^\mu \left[ U^\dagger U \left( \begin{matrix} \nu^{}_{1}
\cr \nu^{}_{2} \cr \nu^{}_{3} \cr\end{matrix}
\right)^{}_{\hspace{-0.08cm} \rm L}
+ U^\dagger R \left(\begin{matrix} N^{}_4 \cr N^{}_5 \cr N^{}_6
\cr\end{matrix}\right)^{}_{\hspace{-0.08cm} \rm L}
+ U^\dagger R^\prime \left(\begin{matrix} N^{\prime}_7 \cr N^{\prime}_8 \cr
N^{\prime}_9 \cr\end{matrix}\right)^{}_{\hspace{-0.08cm} \rm L}
\hspace{0.05cm} \right] Z^0_\mu \right. \hspace{0.4cm}
\nonumber \\
\hspace{-0.2cm} & & \hspace{-0.2cm}
+ \left. \overline{\big(\begin{matrix} N^{}_4 & N^{}_5 & N^{}_6
\end{matrix}\big)^{}_{\rm L}}
\hspace{0.1cm} \gamma^\mu \left[ R^\dagger U \left( \begin{matrix} \nu^{}_{1}
\cr \nu^{}_{2} \cr \nu^{}_{3} \cr\end{matrix}
\right)^{}_{\hspace{-0.08cm} \rm L}
+ R^\dagger R \left( \begin{matrix} N^{}_{4}
\cr N^{}_{5} \cr N^{}_{6} \cr\end{matrix}
\right)^{}_{\hspace{-0.08cm} \rm L}
+ R^\dagger R^\prime \left(\begin{matrix} N^{\prime}_7 \cr N^{\prime}_8 \cr
N^{\prime}_9 \cr\end{matrix}\right)^{}_{\hspace{-0.08cm} \rm L}
\hspace{0.05cm} \right] Z^0_\mu \right.
\nonumber \\
\hspace{-0.2cm} & & \hspace{-0.2cm}
+ \left. \overline{\big(\begin{matrix} N^{\prime}_7 & N^{\prime}_8 &
N^{\prime}_9 \end{matrix}\big)^{}_{\rm L}}
\hspace{0.1cm} \gamma^\mu \left[ R^{\prime\dagger} U \left( \begin{matrix}
\nu^{}_{1} \cr \nu^{}_{2} \cr \nu^{}_{3} \cr\end{matrix}
\right)^{}_{\hspace{-0.08cm} \rm L}
+ R^{\prime\dagger} R \left( \begin{matrix} N^{}_{4}
\cr N^{}_{5} \cr N^{}_{6} \cr\end{matrix}
\right)^{}_{\hspace{-0.08cm} \rm L}
+ R^{\prime\dagger} R^\prime \left(\begin{matrix} N^{\prime}_7 \cr
N^{\prime}_8 \cr N^{\prime}_9 \cr\end{matrix}\right)^{}_{\hspace{-0.08cm} \rm L}
\hspace{0.05cm} \right] Z^0_\mu \hspace{0.05cm} \right\} \; ,
\label{12}
\end{eqnarray}
where $U \equiv A^{}_2 A^{}_1 U^{}_0$ is the effective $3 \times 3$
PMNS matrix, $R \equiv A^{}_2 R^{}_1$ and $R^\prime \equiv R^{}_2$
stand respectively for the contributions of heavy neutrinos and extra
neutral fermions to ${\cal L}^{\prime}_{\rm cc}$ and
${\cal L}^{\prime}_{\rm nc}$. The unitarity of $\mathbb{U}^\prime$
assures a correlation among $U$, $R$ and $R^\prime$:
\begin{eqnarray}
U U^{\dagger} = \left(A^{}_2 A^{}_1\right) \left(A^{}_2 A^{}_1\right)^\dagger
= I - R R^{\dagger} - R^{\prime} R^{\prime \dagger} \; ,
\label{13}
\end{eqnarray}
implying that the PMNS matrix $U$ slightly deviates from the unitary
matrix $U^{}_0$. Note that both $A^{}_1$ and $A^{}_2$ are the lower
triangular matrices, and hence their product $A^{}_2 A^{}_1$ is also
a lower triangular matrix. As three of the nine phases in $A^{}_2$
and $R^{}_2$ can be rotated away from redefining the phases of three
charged-lepton fields, we are totally left with thirty-nine original
flavor parameters in the canonical seesaw mechanism: three heavy Majorana
neutrino masses, three extra neutral fermion masses, eighteen rotation
angles and fifteen independent CP-violating phases. Note, however, that
the vanishing $(2,2)$, $(1,3)$ and $(3,1)$ entries of the $9 \times 9$
mass matrix in Eq.~(\ref{9}) may offer twelve constraint conditions on
the parameter space of the inverse seesaw mechanism, leading us to
 twenty-seven independent seesaw flavor parameter at last.

On the other hand, $U$, $R$ and $R^\prime$ are {\it inherently} correlated
with one another through the exact seesaw relationship in the physical
(mass) basis of all the nine Majorana fermions
\footnote{In comparison, the {\it approximate} inverse seesaw formula in
a general flavor basis is known as $M^{}_\nu \equiv U^{}_0 D^{}_\nu U^T_0
\simeq M^{}_{\rm D} \left(M^T_S\right)^{-1} \mu
\left(M^{}_S\right)^{-1} M^T_{\rm D}$, where one may assume the mass scale
of $\mu$ to be considerably smaller than the Fermi scale
(for example, around the keV scale). In addition, the mass scale of
$M^{}_{\rm D}$ may be remarkably smaller than that of $M^{}_S$, making
a suppression of the mass scale of $M^{}_\nu$ down to ${\cal O}(0.1)$ eV
much easier than the canonical seesaw mechanism even if the heavy degrees
of freedom are located in the TeV regions.},
\begin{eqnarray}
U D^{}_\nu U^T = \left({\rm i} R\right) D^{}_N \left({\rm i} R\right)^T
+ \left({\rm i} R^{\prime}\right) D^{}_S \left({\rm i} R^{\prime}
\right)^T \; ,
\label{14}
\end{eqnarray}
which can easily be achieved from substituting Eq.~(\ref{11}) into
Eq.~(\ref{10}). This elegant result looks like a straightforward
extension of the canonical seesaw relation obtained in Eq.~(\ref{5}).
As the inverse seesaw mechanism aims to significantly lower
the mass scales of those new degrees of freedom added to the SM while
keeping the strengths of relevant Yukawa interactions appreciable,
it is not interesting to consider one or both of the two terms on the
right-hand side of Eq.~(\ref{14}) to be as small as the term on the
left-hand side of Eq.~(\ref{14}). Instead, one is more interested in
the case that both the mass scales of $D^{}_N$ and $D^{}_S$ are close
to the experimentally accessible TeV scale, and both the magnitudes of
$R$ and $R^\prime$ are not strongly suppressed. Then a fined-tuned
cancellation between the contribution from heavy Majorana neutrinos
and that from extra neutral fermions is very likely and even unavoidable.
Whether such a structural cancellation is acceptable or not depends on
the naturalness criterion to be set for this issue.

No matter how natural a specific inverse seesaw model working in the
TeV region can be to generate tiny masses for the three active Majorana
neutrinos, its prediction for their large flavor mixing angles is
sensitive to the details of the seesaw flavor textures. But the smallness
of all the active-sterile flavor mixing angles appearing in $R$ and $R^\prime$
are well constrained thanks to the experimental bounds on non-unitarity
of the $3\times 3$ PMNS matrix $U$~\cite{Blennow:2023mqx,Xing:2024gmy} (see
also Refs.~\cite{Antusch:2006vwa,Antusch:2009gn,Blennow:2016jkn,Hu:2020oba,
Wang:2021rsi} for some earlier works in this connection, and
Ref.~\cite{KM3NeT:2025ftj} for the latest experimental constraint
obtained by the KM3NeT Collaboration), and therefore the inverse seesaw
framework is also likely to act as a {\it cross seesaw} system as
illustrated in Fig.~\ref{fig1}.

\section{Further remarks}

The canonical and inverse seesaw mechanisms have been extensively studied
in the past decades, but most works have started from the approximate
seesaw formulas in the flavor basis and focused on the specific models
and their phenomenological consequences. In the present paper we have
tried to understand why the masses of three active neutrinos are tiny
and why their flavor mixing effects are significant without loss of any
generality by making use of the exact seesaw relations in the physical
(mass) basis of all the relevant lepton fields. To do so, we have sought
help from a complete Euler-like block parametrization of the seesaw flavor
structure for either of the two mechanisms.
We find that the observed flavor mixing pattern of three active neutrinos
should be an emergent consequence of the canonical seesaw mechanism, in
which case one is naturally left with a novel cross seesaw framework. We
have also pointed out that the inverse seesaw mechanism takes effect under
the condition of a fine-tuned cancellation between its two sets of new
degrees of freedom, in which case the large active flavor mixing effects
are an emergent phenomenon as well.

It is certainly difficult to draw more general and interesting conclusions
from such model-independent discussions, but we believe that the exact seesaw
relations in the mass basis deserve more attention in the era of precision
measurements. The reason is simply that we are strongly motivated to test
and even pin down the most reasonable mechanisms of neutrino mass generation
and lepton flavor mixing by calculating all the possible observable quantities
in terms of the original seesaw flavor parameters (as done in
Refs.~\cite{Xing:2024xwb,Xing:2024gmy}), so as to find a correct path to
go beyond the SM and finally arrive at a UV-complete theory.

Let us end with Table~\ref{table1} to compare between the approximate and
exact seesaw relations in the respective flavor and mass bases of the
canonical and inverse seesaw mechanisms. One can see that introducing
more new degrees of freedom to the SM makes the puzzle of why there is
relatively large flavor mixing in the active neutrino sector much more
mysterious, in which case the largeness of three active flavor mixing
angles looks more like an emergent phenomenon.
\begin{table}[t]
\caption{A brief comparison between the approximate seesaw formulas in the flavor
basis and the exact seesaw relations in the mass basis for the canonical and
inverse seesaw mechanisms, where $U = A U^{}_0$ (canonical) or
$U = A^{}_2 A^{}_1 U^{}_0$ together with $R = A^{}_2 R^{}_1$ and
$R^\prime = R^{}_2$ (inverse).}
\label{table1}
\vspace{-0.1cm}
\begin{center}
\begin{tabular}{l|l|l}
\hline\hline
& Canonical seesaw
& Inverse seesaw
\\ \hline
flavor basis
& $M^{}_\nu \equiv U^{}_0 D^{}_\nu U^T_0 \simeq - M^{}_{\rm D} M^{-1}_{\rm R}
M^T_{\rm D}$
& $M^{}_\nu \equiv U^{}_0 D^{}_\nu U^T_0 \simeq M^{}_{\rm D} \left(M^T_S\right)^{-1}
\mu \left(M^{}_S\right)^{-1} M^T_{\rm D}$
\\ \hline
mass basis
& $U D^{}_\nu U^T = \left({\rm i} R\right) D^{}_N \left({\rm i} R\right)^T$
& $U D^{}_\nu U^T = \left({\rm i} R\right) D^{}_N \left({\rm i} R\right)^T
+ \left({\rm i} R^{\prime}\right) D^{}_S \left({\rm i} R^{\prime}
\right)^T$
\\ \hline\hline
\end{tabular}
\end{center}
\end{table}

\section*{Acknowledgements}

The author is greatly indebted to Shun Zhou for many stimulating 
and enlightening discussions, especially for his clarifying the
number of the original inverse seesaw flavor parameters in
several different ways. This research was supported in part by the 
National Natural Science Foundation of China under grant 
No. 12075254.

\newpage

\renewcommand{\theequation}{\thesection.\arabic{equation}}
\section*{Appendix}
\appendix

\setcounter{equation}{0}
\section{The expressions of $A$, $R$ and $U^{}_0$}
\label{A}

Given the Euler-like block parametrization of the $6\times 6$ unitary flavor
mixing matrix $\mathbb{U}$ as decomposed in Eq.~(\ref{3}), the $3\times 3$
active-sterile flavor mixing matrices $A$ and $R$ are found to depend on the
same nine rotation angles $\theta^{}_{ij}$ and nine phase angles $\delta^{}_{ij}$
(for $i = 1, 2, 3$ and $j = 4, 5, 6$). Their explicit and exact expressions
are~\cite{Xing:2007zj,Xing:2011ur},
\begin{eqnarray}
A \hspace{-0.2cm} & = & \hspace{-0.2cm}
\left( \begin{matrix} c^{}_{14} c^{}_{15} c^{}_{16} & 0 & 0
\cr \vspace{-0.45cm} \cr
\begin{array}{l} -c^{}_{14} c^{}_{15} \hat{s}^{}_{16} \hat{s}^*_{26} -
c^{}_{14} \hat{s}^{}_{15} \hat{s}^*_{25} c^{}_{26} \\
-\hat{s}^{}_{14} \hat{s}^*_{24} c^{}_{25} c^{}_{26} \end{array} &
c^{}_{24} c^{}_{25} c^{}_{26} & 0 \cr \vspace{-0.45cm} \cr
\begin{array}{l} -c^{}_{14} c^{}_{15} \hat{s}^{}_{16} c^{}_{26} \hat{s}^*_{36}
+ c^{}_{14} \hat{s}^{}_{15} \hat{s}^*_{25} \hat{s}^{}_{26} \hat{s}^*_{36} \\
- c^{}_{14} \hat{s}^{}_{15} c^{}_{25} \hat{s}^*_{35} c^{}_{36} +
\hat{s}^{}_{14} \hat{s}^*_{24} c^{}_{25} \hat{s}^{}_{26}
\hat{s}^*_{36} \\
+ \hat{s}^{}_{14} \hat{s}^*_{24} \hat{s}^{}_{25} \hat{s}^*_{35}
c^{}_{36} - \hat{s}^{}_{14} c^{}_{24} \hat{s}^*_{34} c^{}_{35}
c^{}_{36} \end{array} &
\begin{array}{l} -c^{}_{24} c^{}_{25} \hat{s}^{}_{26} \hat{s}^*_{36} -
c^{}_{24} \hat{s}^{}_{25} \hat{s}^*_{35} c^{}_{36} \\
-\hat{s}^{}_{24} \hat{s}^*_{34} c^{}_{35} c^{}_{36} \end{array} &
c^{}_{34} c^{}_{35} c^{}_{36} \cr \end{matrix} \right) \; ,
\nonumber \\
R \hspace{-0.2cm} & = & \hspace{-0.2cm}
\left( \begin{matrix} \hat{s}^*_{14} c^{}_{15} c^{}_{16} &
\hat{s}^*_{15} c^{}_{16} & \hat{s}^*_{16} \cr \vspace{-0.45cm} \cr
\begin{array}{l} -\hat{s}^*_{14} c^{}_{15} \hat{s}^{}_{16} \hat{s}^*_{26} -
\hat{s}^*_{14} \hat{s}^{}_{15} \hat{s}^*_{25} c^{}_{26} \\
+ c^{}_{14} \hat{s}^*_{24} c^{}_{25} c^{}_{26} \end{array} & -
\hat{s}^*_{15} \hat{s}^{}_{16} \hat{s}^*_{26} + c^{}_{15}
\hat{s}^*_{25} c^{}_{26} & c^{}_{16} \hat{s}^*_{26} \cr \vspace{-0.45cm} \cr
\begin{array}{l} -\hat{s}^*_{14} c^{}_{15} \hat{s}^{}_{16} c^{}_{26}
\hat{s}^*_{36} + \hat{s}^*_{14} \hat{s}^{}_{15} \hat{s}^*_{25}
\hat{s}^{}_{26} \hat{s}^*_{36} \\ - \hat{s}^*_{14} \hat{s}^{}_{15}
c^{}_{25} \hat{s}^*_{35} c^{}_{36} - c^{}_{14} \hat{s}^*_{24}
c^{}_{25} \hat{s}^{}_{26}
\hat{s}^*_{36} \\
- c^{}_{14} \hat{s}^*_{24} \hat{s}^{}_{25} \hat{s}^*_{35}
c^{}_{36} + c^{}_{14} c^{}_{24} \hat{s}^*_{34} c^{}_{35} c^{}_{36}
\end{array} &
\begin{array}{l} -\hat{s}^*_{15} \hat{s}^{}_{16} c^{}_{26} \hat{s}^*_{36}
- c^{}_{15} \hat{s}^*_{25} \hat{s}^{}_{26} \hat{s}^*_{36} \\
+c^{}_{15} c^{}_{25} \hat{s}^*_{35} c^{}_{36} \end{array} &
c^{}_{16} c^{}_{26} \hat{s}^*_{36} \cr \end{matrix} \right) \; .
\hspace{0.5cm}
\label{A.1}
\end{eqnarray}
Note that three of the nine phases (or their combinations) in $A$ and $R$
can be rotated away through a redefinition of the phases of three
charged-lepton fields, as one may see from ${\cal L}^{}_{\rm cc}$ in
Eq.~(\ref{4}). On the other hand, the $3\times 3$ unitary matrix $U^{}_0$
is of the form
\begin{eqnarray}
U^{}_0 \hspace{-0.2cm} & = & \hspace{-0.2cm}
\left( \begin{matrix} c^{}_{12} c^{}_{13} & \hat{s}^*_{12}
c^{}_{13} & \hat{s}^*_{13} \cr
-\hat{s}^{}_{12} c^{}_{23} -
c^{}_{12} \hat{s}^{}_{13} \hat{s}^*_{23} & c^{}_{12} c^{}_{23} -
\hat{s}^*_{12} \hat{s}^{}_{13} \hat{s}^*_{23} & c^{}_{13}
\hat{s}^*_{23} \cr
\hat{s}^{}_{12} \hat{s}^{}_{23} - c^{}_{12}
\hat{s}^{}_{13} c^{}_{23} & -c^{}_{12} \hat{s}^{}_{23} -
\hat{s}^*_{12} \hat{s}^{}_{13} c^{}_{23} & c^{}_{13} c^{}_{23}
\cr \end{matrix} \right) \; , \hspace{0.4cm}
\label{A.2}
\end{eqnarray}
in which case it is the phase combination $\delta \equiv \delta^{}_{13}
- \delta^{}_{12} - \delta^{}_{23}$ that determines the strength of CP
violation in neutrino oscillations. It is worth remarking that all the
six flavor parameters in $U^{}_0$ are {\it derivational} in the sense
that they can all be derived from the {\it original} seesaw flavor
parameters (the three heavy Majorana neutrino masses in $D^{}_N$, the
nine active-sterile flavor mixing angles and six independent CP-violating
phases in $A$ and $R$)~\cite{Xing:2024xwb,Xing:2024gmy}.

\end{document}